\documentclass[prl,twocolumn,showpacs,floatfix,amsmath,amsfonts]{revtex4}
\usepackage{amsmath}
\usepackage{epsfig}
\usepackage{fancybox}

\begin{document}
%\fancyput(-0.7cm,1cm){\epsfig{file=arriba.eps,width=18.5cm}}

\title{Stabilized vortex solitons in layered Kerr media}

\author{Gaspar D. \surname{Montesinos}}
\affiliation{Departamento de Matem\'aticas, Escuela T\'ecnica
Superior de Ingenieros Industriales, \\
Universidad de Castilla-La Mancha, 13071 Ciudad Real, Spain}

\author{V\'{\i}ctor M. \surname{P\'erez-Garc\'{\i}a}}
\affiliation{Departamento de Matem\'aticas, Escuela T\'ecnica
Superior de Ingenieros Industriales, \\
Universidad de Castilla-La Mancha, 13071 Ciudad Real, Spain}

\author{Humberto \surname{Michinel}}
\affiliation{\'Area de \'Optica, Facultade de Ciencias de Ourense,\\
Universidade de Vigo, As Lagoas s/n, Ourense, ES-32005 Spain.}

\author{Jos\'e R. \surname{Salgueiro}}
\affiliation{\'Area de \'Optica, Facultade de Ciencias de Ourense,\\
Universidade de Vigo, As Lagoas s/n, Ourense, ES-32005 Spain.}

\date{\today}

%%%%%%%%%%%%%%%%%%% ABSTRACT & PACS %%%%%%%%%%%%%%%%%%%%%%%%%%%

\begin{abstract}
In this letter we demonstrate the possibility of stabilizing beams
with angular momentum propagating in Kerr media. Large propagation
distances without filamentation can be achieved in layered media
with alternating focusing and defocusing nonlinearities. Stronger
stabilization can be obtained with the addition of an incoherent
beam. \end{abstract}

% 42.65.Tg: Optical solitons; nonlinear guided waves
% 03.75.Fi Phase coherent atomic ensemble (Bose condensation)

%%%%%%%%%%%%%%%%%%%%%%%%%%%%%%%%%%%%%%%%%%%%%%%%%%%%%%%%%%%%%%%

\maketitle
%\fancyput*(-0.7cm,-25cm){\epsfig{file=abajo.eps,width=18.5cm}}

\emph{Introduction.-} Vortices have been a source
of fascination since the works of Empedocles, Aristotle and Descartes,
who tried to explain the formation of the Earth, its gravity and the
dynamics of the solar system as due to primordial cosmic
vortices.  Many interesting problems related to vortices are open
in different fields such as:
fluid mechanics, high $T_c$ superconductivity,  superfluidity, light
propagation, Bose-Einstein condensation (BEC), cosmology,
biosciences, or solid state physics
\cite{Lug95,Pis99,Sols,experimental}.

In wave mechanics a vortex is a screw phase
dislocation, or defect \cite{nye74}, where the amplitude of the
field vanishes. The phase around the singularity has an integer
number of windings, $\ell$, which plays the role of an angular
momentum. For fields with non-vanishing boundary conditions, this
number is a conserved quantity and governs the interactions
between vortices as if they were endowed with electrostatic
charges. Thus, $\ell$ is usually called the {}``topological
charge{}'' of the defect.

In  Optics there has been a strong interest on the
so called ``vortex solitons", i. e. robust distributions of light
of vortex type in which nonlinearity could compensate diffraction
leading to stationary propagation. However, in self-focusing
Kerr media, a finite size beam containing a vortex always
 destabilizes and forms a filamentary structure \cite{kruglov85}.
This also stands for saturable self-focusing nonlinearities
\cite{firth97}. Vortex solitons have been studied in many other different
optical systems (see e.g. the review \cite{Yuri}) and in most realistic
cases they tend to be unstable.

In  this paper we propose to use layered Kerr media,
which are self-focusing in average, to obtain
stable propagation of vortex solitons up to very long
distances. Our ideas are also extended to the field of matter waves.

\emph{Stabilized solitons.-} In Ref.  \cite{Berge} an idea to prevent
collapse was proposed based on the modulation of the Kerr coefficient of an
optical material along the propagation direction. In that way, a propagating beam
focuses and expands in alternating regions and becomes stabilized
in average. This idea has been further explored in
\cite{IsaacLayered,Ueda,BM,Gaspar,IMACS,PROLO}.

The propagation of a paraxial monochromatic beam in a Kerr medium
is modeled by equations of the type (in adimensional units)
\begin{equation}
\label{NLS} i \frac{\partial u}{\partial z}  = - \frac{1}{2}
\Delta u + g(z)  |u|^2 u,
 \end{equation}
where $u(x,y,z): \mathbb{R}^2\times \mathbb{R}^+\rightarrow
\mathbb{C}$  is the slowly varying amplitude of the beam envelope,
$\Delta =
\partial^2/\partial x^2 + \partial^2/\partial y^2,$  and $g(z)$ is
a periodic function accounting for the modulation of the
nonlinearity. It is well known that, if $g$ is
constant, there is a stationary radially symmetric solution of Eq.
\eqref{NLS} (the so-called Townes soliton): $u({\mathbf r},
z) = \Phi_0(r)e^{i\lambda z}$. This solution is {\em unstable}
 since any generic slight
perturbation of the initial condition will yield either collapse
or spread of the distribution. In previous works, it has been
shown that the structure which arises when the nonlinearity is
modulated as described above is a stabilized Townes soliton (STS)
\cite{Gaspar,IMACS}.

\begin{figure}
\begin{center}
\epsfig{file=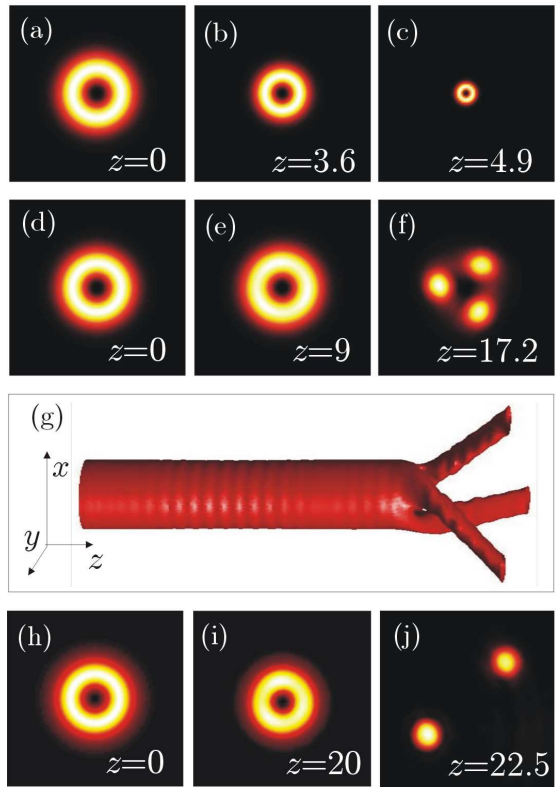,width=0.85\columnwidth}
 \caption{[Color online] Evolution of initial data of vortex type obtained as a
stationary solution of Eq. (\ref{NLS})  for $g_v = -24.15$ under
different parameter variations. (a-c) Evolution for constant $g =
-8\pi$. (d-f) Evolution with modulated nonlinearity $g(z) = -8\pi
+20 \pi \cos 40 z$. (g) Isosurface plot of $u(x,y,z)$ of the same
simulation as in (d-f) showing the development of the instability.
(h-j) Evolution with modulated nonlinearity $g(z) = -27 + 20 \pi
\cos 40 z$.\label{prima}}\end{center}
\end{figure}

\emph{Partially stabilized vortex solitons.-} However, Townes
solitons are not the only stationary solutions of Eq. \eqref{NLS}
for constant $g$. There also exist vortex-type solutions of the
form $u({\mathbf r}, z) = \Phi_{\ell}(r)e^{i\ell
\theta}e^{i\lambda z}$ which are unstable as well. Therefore, one
could naively expect that the same stabilization mechanism
proposed in Ref. \cite{Berge} could be applied to achieve
stabilization of  these solutions, i.e. to induce alternative
expanding and squeezing cycles of the vortex by a periodic
modulation of the nonlinear coefficient. To study further this
possibility we have computed numerically the profiles of vortex
solutions of Eq. \eqref{NLS} corresponding to a constant value of
$g_{v}= -24.15$ (the critical one for vortex solutions), by using
a standard shooting method. We have then studied the evolution of
this stationary solution numerically using a special
pseudospectral method described in Ref. \cite{IMACS} in different
situations. Obviously if we compute the evolution of this vortex
solution for constant $g>g_{v}$ we would have expansion while for
$g=g_0<g_{v}$ it would collapse. In Fig. \ref{prima}(a,b,c) we
plot the collapse of a vortex with supercritical value of $g_0 =
-8\pi$, while in Fig. \ref{prima}(d,e,f) we present some snapshots
of the evolution of the vortex after the addition of a stabilizing
term to the nonlinearity $g(z) = g_0 + g_1 \cos \Omega z$ for $g_0
= -8\pi, g_1 = 20\pi, \Omega = 40$. We can see how the periodic
modulation of the nonlinearity retards the filamentation of the
vortex, which now propagates for a longer distance before breaking
into several stabilized solitons.
The splitting length and number of outgoing stabilized solitons
depend on the specific parameters of the nonlinearity. For
instance choosing a different set of parameters $g_0 = -27, g_1 =
20\pi, \Omega = 40$ increases the stability of the vortex and
decreases the number of filaments. Since each emerging beam is
close to an STS the excess energy is eliminated in the form of
radiation which is removed by the absorbing boundary conditions of
our numerical scheme. In this paper we have chosen a smooth form
for the modulation $g(z)$ but similar results are obtained when
$g(z)$ is taking as piecewise constant.

We have made an extensive search in the parameter space for
modulations of the form $g(z) = g_0 + g_1 \cos \Omega z$ and we
have not been able to find any parameter combination allowing
indefinite stabilization of the vortex soliton. Concerning
finite-dimensional reduced models for the evolution of the
effective width of the solutions, such as those successfully used
for nodeless beams in  Refs. \cite{IsaacLayered,Ueda,BM,Gaspar},
we must stress that as pointed out in Ref. \cite{IsaacLayered}
these formulations do not reflect correctly the dynamics and
instabilities of vortex solutions.

\emph{Vector systems.-} From the previous analysis it seems that a
vortex can only be \emph{partially} stabilized in the framework of
Eq. (\ref{NLS}), i.e. in scalar systems. Recent works point out
the important fact that the incoherent interaction of two
components could provide, in saturable media, an effective
waveguide for the vortex, leading to a more stable behavior
\cite{Segev}. Following this idea we consider now a vector
two-component system with Kerr interactions of the form:
\begin{subequations}
\label{Manakoveqs}\begin{eqnarray}
i \frac{\partial u_1}{\partial z}  & = &
- \frac{1}{2} \Delta u_1 + g(z) \left(a_{11} |u_1|^2 +a_{12} |u_2|^2\right)u_1, \\
i \frac{\partial u_2}{\partial z}  & = & - \frac{1}{2} \Delta u_2
+ g(z) \left(a_{21} |u_1|^2 +a_{22}|u_2|^2\right)u_2,
\end{eqnarray}
\end{subequations}
where $a_{jk} \in \mathbb{R}$ are the nonlinear coupling
coefficients and $g(z)$ accounts for the  modulation of
the nonlinearity. We will denote $I_j =
\int_{\mathbb{R}^2} |u_j|^2 dx dy$. Although this system is
conservative, in the numerical simulations to be shown later we
incorporate absorbing boundary conditions in order to get rid of
the radiation. Therefore, in practice, there will be a decrease of
$I_j$ during the propagation.

Eqs. (\ref{Manakoveqs}) are an extension of the Manakov system
\cite{Manakov} to two transverse dimensions. Among other
situations these equations model the propagation of two circularly
polarized beams with opposite polarizations leading to specific
factors $a_{11} = a_{22} = 1, a_{12} = a_{21} = 2$.
In the context of BEC, these equations (with an additional
trapping term) describe the dynamics of multicomponent
quasi-two dimensional condensates, $u_{j}$ being the wavefunctions
for the atomic species involved. The formation of vector solitons
composed of appropriate fractions of Townes states has been studied
in Ref. \cite{PROLO}.

Our idea is to choose $g(z)$ to get the Townes soliton $u_1$ stabilized.
As the coupling terms in Eq. (\ref{Manakoveqs}) would
provide an effective waveguide for $u_2$, it seems reasonable
that, when $I_2 \ll I_1$, the guiding effect will dominate
over self-interaction and the vortex could become stabilized.

\emph{Limit of small $u_2$.-} Let us first consider the case
of constant $g$ and $I_2 \ll I_1$ so that Eqs. (\ref{Manakoveqs}) become:
\begin{subequations}
\label{fulla}
\begin{eqnarray}
i \frac{\partial u_1}{\partial z}   & \simeq &
- \frac{1}{2} \Delta u_1 + g a_{11}|u_1|^2 u_1, \label{approximatea} \\
\label{approximateb} i \frac{\partial u_2}{\partial z}  & \simeq &
\left(- \frac{1}{2} \Delta + g a_{21} |u_1|^2\right) u_2.
 \end{eqnarray}
\end{subequations}
Taking $u_1(r,z) = \Phi_0(r)e^{i\lambda_0z}$, then Eq.
\eqref{approximateb} is a linear two-dimensional Schr\"odinger
problem in which the role of the potential is played by
$|\Phi_0|^2$. Following Ref. \cite{estimaciones} we can  bound the
number of $\ell$-wave bound states in this potential, by $N_{\ell}
<  (a_{12}g/\ell) \int_{\mathbb{R}^+} r |u_1|^2 dr$ . Vortex-type
solutions with smallest topological charge are those with $\ell =
1$. Thus, taking into account that for a Townes soliton $gI_1 =
0.931$ we get $N_{\ell=1} <  0.931 a_{12}$. Therefore, we can
expect the existence of a unique $\ell = 1$ stationary vortex
solution of Eq. (\ref{approximateb}) in the case $a_{21} = 2$.
This will be denoted hereafter as $u_2 =
V(r)e^{i\theta}e^{i\lambda_vz}$. Let us notice that for $\ell \geq
2$  we get always $N_{\ell \geq 2} < 1$ thus ruling out the
possibility of obtaining higher order vortices. We have
numerically found the profile of this vortex solution $V(r)$ by
using a standard shooting method.

\begin{figure}
\hspace{3cm} \epsfig{file=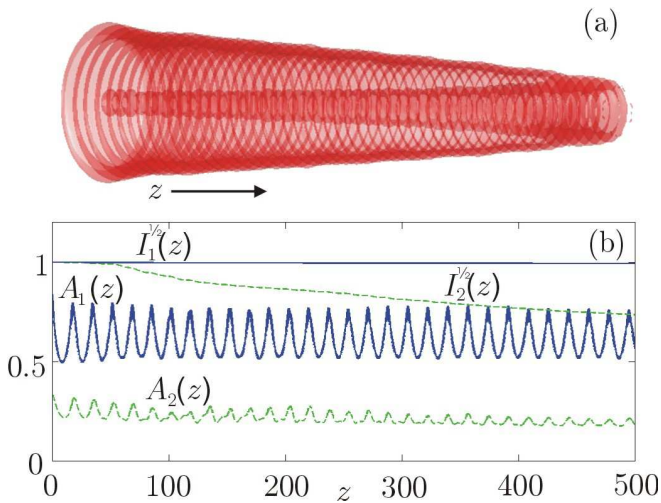,width=\columnwidth}
\caption{[Color online] Solutions of Eqs. (\ref{Manakoveqs}) for
initial data $u_1(r,0)= \Phi_0(r), u_2 = \alpha V(r)e^{i\theta}$
with $\alpha = 0.1$ in a grid of $810 \times 810$ points on the
spatial region [-40,40] showing stable (but dissipative due to the
effect of radiation) propagation of the vortex in  the full range
$z\in[0,500]$. (a) Isosurface plot of $u_2(x,y,z)$ spanning all
the propagation range. (b) Evolution of the norms $I_1^{1/2}(z),
I_2^{1/2}(z)$ and of the amplitudes $A_1(z) = \max_{(x,y)} |u_1|$,
$A_2(z) = \max_{(x,y)} |u_2|$ of both components. \label{stable}}
\end{figure}

Choosing an appropriate modulation for $g$ allows us to stabilize
$u_1$. For Eq. (\ref{approximateb}),  the potential --the
stabilized Townes soliton-- oscillates with a fast frequency of
the order of $\Omega$ and another slower one of dynamical origin
\cite{Gaspar}. In this linear case we can apply to $u_2$ the quantum
mechanical theory of fast perturbations \cite{Galindo} to account
for the effect of the fast modulation on the vortex. The main
result of this theory is that the vortex will remain unaffected by
the fast perturbation in the potential provided the modulation
period $T$ satisfies $\Delta_{u_2} \bar{H} \ll 1/T$. In
our case  $\bar{H} = \frac{1}{T} \int_0^T \left[-\frac{1}{2}\Delta
+ g(t)a_{21}|u_{1}|^2 \right]$ and this inequality imposes $T \ll
1.12$ which requires $\Omega \gg 5.6$. The STS oscillation
of lower frequency will induce the same modulation in the vortex.
\begin{figure}
\epsfig{file=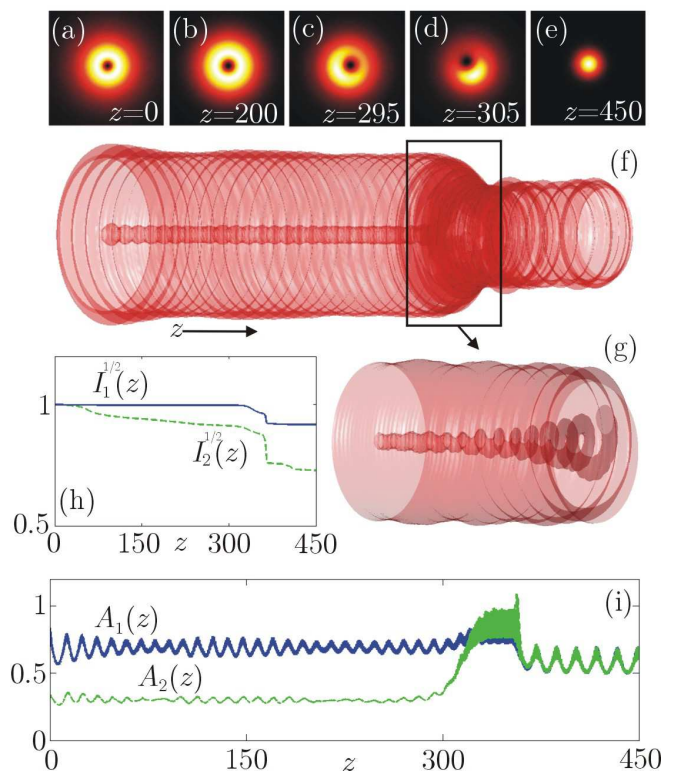,width=\columnwidth} \caption{[Color
online] Same as Fig. \ref{stable} for $\alpha = 0.32$. (a)-(e)
Pseudocolor plots of $u_2(x,y,z)$ for $z=0, 200, 295, 305, 450$.
(f) Isosurface plot of $u_2(x,y,z)$ spanning all the propagation
range. (g) Details of the region in which the vortex spirals out
of $u_2$ for $z\in [280,319]$. (h) Evolution of the norms of both
components $I_1^{1/2}(z), I_2^{1/2}(z)$ showing the readjustment
of the norms after the vortex is ejected and a stabilized vector
soliton is formed. (i) Amplitudes $A_1(z) = \max_{(x,y)} |u_1|$
and $A_2(z) = \max_{(x,y)} |u_2|$ of both components.
\label{tertia}}
\end{figure}
To verify our previous considerations we have simulated Eqs.
(\ref{fulla}) with initial conditions $u_{1}(r,0) = \Phi_0(r),
u_2(r,\theta,0) =  V(r)e^{i\theta}$ and $g(z) = -2\pi + 8\pi \cos
40 z$. Full stabilization of the vortex is observed, its
oscillations following the pattern predicted above, i.e. there is only a residual
 fast oscillation in the vortex component and its slow oscillation follows
that of the stabilized Townes soliton in $u_1$.

\emph{Fully nonlinear regime.-} To account for the case of finite
$u_2$ we have studied numerically the solutions of  Eqs.
(\ref{Manakoveqs}) taking as initial data $u_1(r,0) = \Phi_0(r),
u_2(r,\theta,0) = \alpha V(r) e^{i\theta}$ and $g(z) = -2\pi + 8\pi
\cos 40 z$. We have repeated most simulations starting with the full stationary solutions
of Eqs. (\ref{Manakoveqs}) and found similar results.

For small values of $\alpha$ (e.g. $\alpha = 0.1$ as in Fig.
\ref{stable}) the vortex is fully stabilized up to the maximum
propagation distances studied. However, a continuous loss of
energy is observed during propagation. We think that this power
damping is due to radiation emitted by the vortex and it is
related to the continuous background oscillations of the
stabilizing Townes beam. We have tested these numerical
simulations both by the spectral numerical scheme of Ref.
\cite{IMACS}, for which radiation is eliminated by means of an
absorbing potential and by a finite-difference Crank-Nicholson
type scheme with transparent boundary conditions with the same
results.

For larger values of $\alpha$ (e.g. $\alpha = 0.32$ as in Fig.
\ref{tertia}), the vortex destabilizes at long propagation
distances due to the effect of nonlinear interactions between the
guiding Townes soliton and the vortex. We can see that, although
the perturbation is not small, the vortex propagates for very long
distances of about 300 propagation units (compare this with the
results shown in Fig. \ref{prima}). In the region of stable
propagation the vortex amplitude slowly decays due to the emission
of radiation until a stabilized vector soliton is formed. In this
process, both components emit radiation to readjust their norms to
satisfy the relation $I_1 + I_2 = I_{Townes}$ \cite{PROLO}. From
Fig. \ref{tertia}(i) it is clear that the oscillations of the
vortex amplitude basically contain only the slower frequency and
that our previous arguments apply here. The addition of noise to
the initial data of 1\% in amplitude triggers the instability
faster but even in that case the vortex propagates for more than
125 adimensional units before the instability sets in.

Finally let us briefly comment that for $\alpha = \mathcal{O}(1)$
there is a different branch of bound states of Eqs.
(\ref{Manakoveqs}) in the fully nonlinear regime, corresponding to
thin Townes solitons localized near the bottom of the vortex
solutions. This soliton act as a pinning potential for the vortex
but it is not enough to stabilize vortex-type solutions.

\emph{Applications to matter waves.-} The previous results have
also implications in the field of matter waves because of the
close analogy of Eqs. (\ref{Manakoveqs}) with the equations of
evolution of a multi-component Bose-Einstein condensate in the
mean field approximation. The analysis of vortices in dilute-gas
BECs has been a very hot topic in the last years, specially after
their experimental generation with different setups
\cite{experimental}. In multicomponent BEC systems the interaction
coefficients $a_{ij}$ are proportional to the respective
scattering lengths and the effective two-dimensionality can be
achieved by confining the condensate tightly along one specific
direction. The condition $N_{1} < 0.931 a_{12}$ has relevance
since it imposes restrictions to the atomic species which can be
used to trap a vortex. For instance the cross-interaction
coefficient in multicomponent condensates made of different
hyperfine levels of $^{87}$Rb does not satisfy this condition.
However, bosonic K-Rb mixtures such as the one described in Refs.
\cite{K-Rb,K-Rb2} could be used because of the large scattering
length of the collisions K-Rb. In this scenario the results
presented here could be extended to tightly confined BEC systems.
Our predictions would imply the existence of self-supported vortex
solitons which could be generated using Feschbach resonance
management techniques.

In conclusion, we have shown stabilization of vortex solitons for
long propagation distances in stratified Kerr media by control of
the nonlinear coefficient and stronger
 stabilization by the use of two
combined beams. Our results can also be used to stabilize vortices
in certain types of multicomponent Bose-Einstein condensates.

This work has been partially supported by Ministerio de
Ciencia y Tecnolog\'{\i}a (Spain) under grants BFM2000-0521,
BFM2003-02832, TIC-2000-1105-C03-01 and by
the Junta de Comunidades de Castilla-La Mancha under grant
PAC-02-002. G. D. M. acknowledges support from grant AP2001-0535
from MECD.

%%%%%%%%%%%%%%%%%%% BIBLIOGRAPHY %%%%%%%%%%%%%%%%%%%%%%%%%%%%

\end{document}